\documentclass[12pt]{article}
\usepackage{graphicx}

\newcommand\pubnumber{KA-TP-36-2017\\MCnet-17-18}
\newcommand\pubdate{\today}

\def\napoli{Institute for Theoretical Physics, Karlsruhe Institute of Technology, Wolfgang-Gaede-Stra\ss e 1, 76131 Karlsruhe, Germany}
\def\support{\footnote{Speaker}}

\def\Title#1{\begin{center} {\Large #1 } \end{center}}
\def\Author#1{\begin{center}{ \sc #1} \end{center}}
\def\Address#1{\begin{center}{ \it #1} \end{center}}

\newcommand\pubblock{\rightline{\begin{tabular}{l} \pubnumber\\
         \pubdate  \end{tabular}}}
\newenvironment{Abstract}{\begin{quotation}  }{\end{quotation}}
\newenvironment{Presented}{\begin{quotation} \begin{center} 
             PRESENTED AT\end{center}\bigskip 
      \begin{center}\begin{large}}{\end{large}\end{center} \end{quotation}}

\def\beq{\begin{equation}}
\def\eeq#1{\label{#1}\end{equation}}
\def\eeqn{\end{equation}}

\def\beqa{\begin{eqnarray}}
\def\eeqa#1{\label{#1}\end{eqnarray}}
\def\eeqan{\end{eqnarray}}

\let\bar=\overbar

\def\Dslash{\not{\hbox{\kern-4pt $D$}}}
\def\dslash{\not{\hbox{\kern-2pt $\del$}}}

\def\msb{{\bar{\ssstyle M \kern -1pt S}}}

\begin{document}
\begin{titlepage}
\pubblock

\vfill
\Title{Soft and diffractive scattering}
\vfill
\Author{Patrick Kirchgae\ss er\support , Stefan Gieseke and Frasher Loshaj}
\Address{\napoli}
\vfill
\begin{Abstract}
A model for soft and diffractive scattering is presented
which is incorporated seamlessly into the pre-existing MPI model
in the Monte Carlo Event Generator Herwig.
With the improved model we are able to describe general aspects of 
Minimum Bias and Underlying Event data.

\end{Abstract}
\vfill
\begin{Presented}
Presented at EDS Blois 2017, Prague, \\ Czech Republic, June 26-30, 2017
\end{Presented}
\vfill
\end{titlepage}
\def\thefootnote{\fnsymbol{footnote}}
\setcounter{footnote}{0}

\section{Introduction}
The correct simulation of soft and diffractive interactions is an essential part
in the physics program of Monte Carlo event generator development. 
It is necessary for the understanding of the Underlying Event (UE) 
and Minimum Bias (MB) measurements. 
Also in the light of pile-up modelling and background 
reduction it plays an important role since every hard 
collision is accompanied by many additional $pp$ collisions.
Furthermore the study of soft physics at low $p_{\perp}$ is a rich and
interesting topic by itself.
This talk explains the current status of the model for soft and diffractive scattering
\cite{Gieseke:2016fpz, Gieseke:2017yfk, Gieseke:2017ccw}
in the Monte Carlo event generator Herwig \,\cite{Bellm:2017bvx,Bellm:2015jjp,Bahr:2008pv}.
For a more detailed description we refer to the references given above.

\section{Model for soft interactions}
\begin{figure}[htb]
\centering
\includegraphics[height=1.5in]{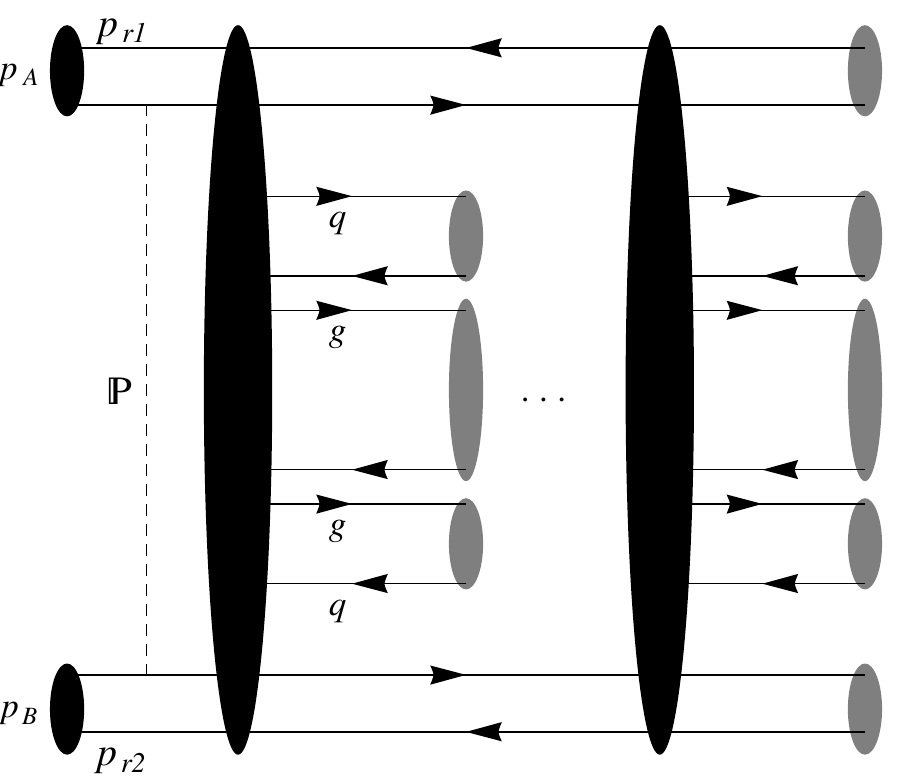}
\caption{Cluster formation in the multiperipheral final state with multiple interactions.}
\label{fig:multiperipheral}
\end{figure}

As explained in \cite{Gieseke:2016fpz} the new model resolves
some of the short comings of the old model.
The number of soft interactions is given by the eikonal approximation
such that 
\begin{equation}
\sigma_{\mathrm{tot}}(s) = \sigma_{\mathrm{soft}}(s) + \sigma_{\mathrm{semi-hard}}(s),
\end{equation}
 where $\sigma_{\mathrm{tot}}(s)$ is given by the Donnachie-Landshoff parametrization\,
\cite{Donnachie:1992ny} for data from the LHC.
The soft interactions are modelled according to the properties of cut-pomerons,
were each cut-pomeron exchange leads to a ladder of partons obeying 
multiperiperipheral kinematics \cite{Baker:1976cv}.
The number of partons in the ladder is sampled from a poissonian distribution with mean at
\begin{equation}
\langle N \rangle \approx N_{\mathrm{ladder}} \times \ln \frac{(p_1 + p_2)^2}{m_{\mathrm{rem}}^2},
\label{eq:N}
\end{equation}
where $p_{1,2}$ denote the four momenta of the proton remnants and 
$m_{\mathrm{rem}}$ is the constituent mass of the proton remnant.
$N_{\mathrm{ladder}}$ is a parameter of the model and a dedicated 
energy dependend tune to MB data at 900 GeV, 7 TeV and 13 TeV showed that 
$N_{\mathrm{ladder}}$ can be parametrized according to the following power law
\begin{equation}
N_{\mathrm{ladder}} = N_{0}\left( \frac{s}{\mathrm{TeV}^2}\right)^{-0.08},
\end{equation}
where $N_{0} \approx 1$.
This light energy dependence of the $N_{\mathrm{ladder}}$ parameter seems to cancel
the $\ln(s)$ term in eq.\,\ref{eq:N} which leads to a constant average number of
partons within the ladder with rising energy. 
It would be interesting to investigate this subject in further projects.
After the number of partons in the ladder is sampled the longitudinal energy fraction
given to each parton in the ladder is calculated such that the partons
in the ladder are distributed equally in the rapidity interval between the two remnants.
A sketch of the parton ladder is shown in fig.\,\ref{fig:multiperipheral} 
for the case of two soft interactions.
Since we aim for a description at parton level one has to take colour connections between
the extracted partons into account. The remnants are by default anti-coloured.
Therefore one needs the first two partons in the ladder to be a quark anti-quark pair 
in order to get the correct colour connections within the ladder.
We look at low-$p_{\perp}$ interactions meaning that the partons in the ladder are treated 
non-perturbatively. They split into quark antiquark pairs and form clusters with the
colour connected quarks. These clusters are then handled by the hadronization model
from Herwig.

\section{Diffraction}

\begin{figure}[htb]
\centering
\includegraphics[scale=0.6]{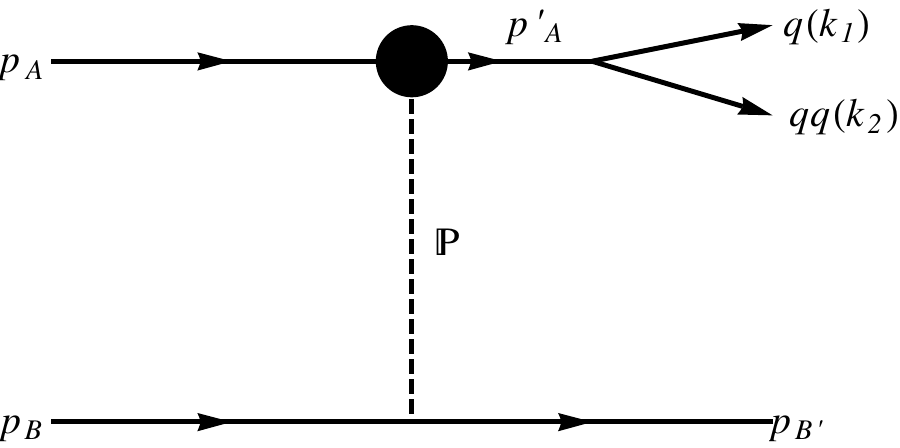}\\
\includegraphics[scale=0.6]{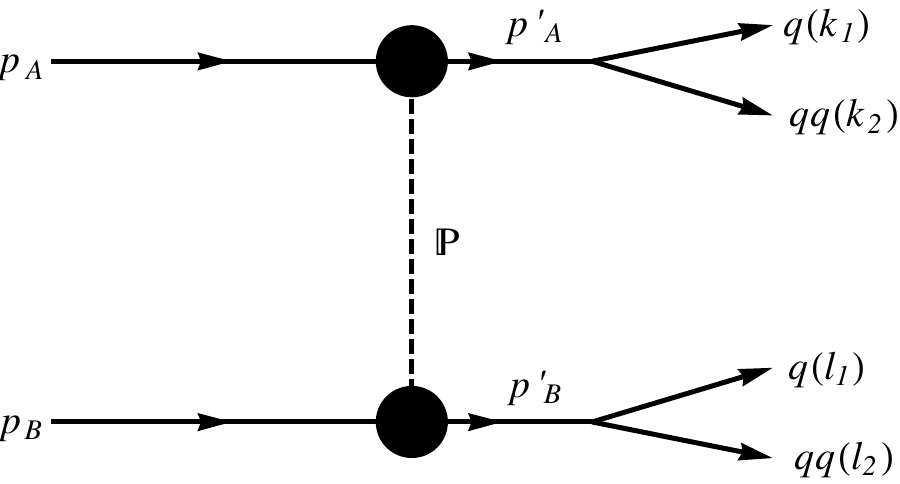}
\caption{Diffractive dissociation for single (top) and double (bottom) diffraction.}
\label{fig:diffraction}
\end{figure}

We implemented a model for soft diffraction which is a completely new feature in Herwig.
This allows us to describe diffractive observables like the rapidity gap in the forward
region $\Delta \eta^F$. The diffractive events are generated according to the differential
cross section for single and double diffraction which can be described by Regge theory 
and the generalized optical theorem. 
The two processes are depicted in fig.\,\ref{fig:diffraction}.
The final states are treated fully non-perturbatively. The quark (q) and diquark (qq) form a 
cluster with diffractive mass $M$ and are stretched along the direction of the dissociated
proton. There is no cross talk between the two diffractive systems which would
lead to a big cluster ranging over the whole rapidity interval.
The mass of the diffractive systems is sampled according to $(1/M^2)^{\lambda}$ where $\lambda$
is the pomeron intercept. 
This leads to a tail of high diffractive masses which is necessary in order 
to simulate events with medium rapidity gaps.
After sampling the invariant mass and the scattering angle the outgoing momenta of the
diffractive systems are constructed.
The dissociated proton is then decayed further into an quark antiquark pair moving collinear
to the dissociated proton.
A cluster is formed out of the quark antiquark pair and then handled by 
the hadronization model.
In order to implement diffraction into the framework of the existing MPI model the simplest
approach was to assume that the cross section for hard and soft interactions only sum up to a
certain fraction of the total cross section. The contribution from diffractive events 
is assumed to be roughly about a rate of 20-25$\%$ where we get the value from data.


\section{Results}
\begin{figure}[htb]
\centering
\includegraphics[scale=0.5]{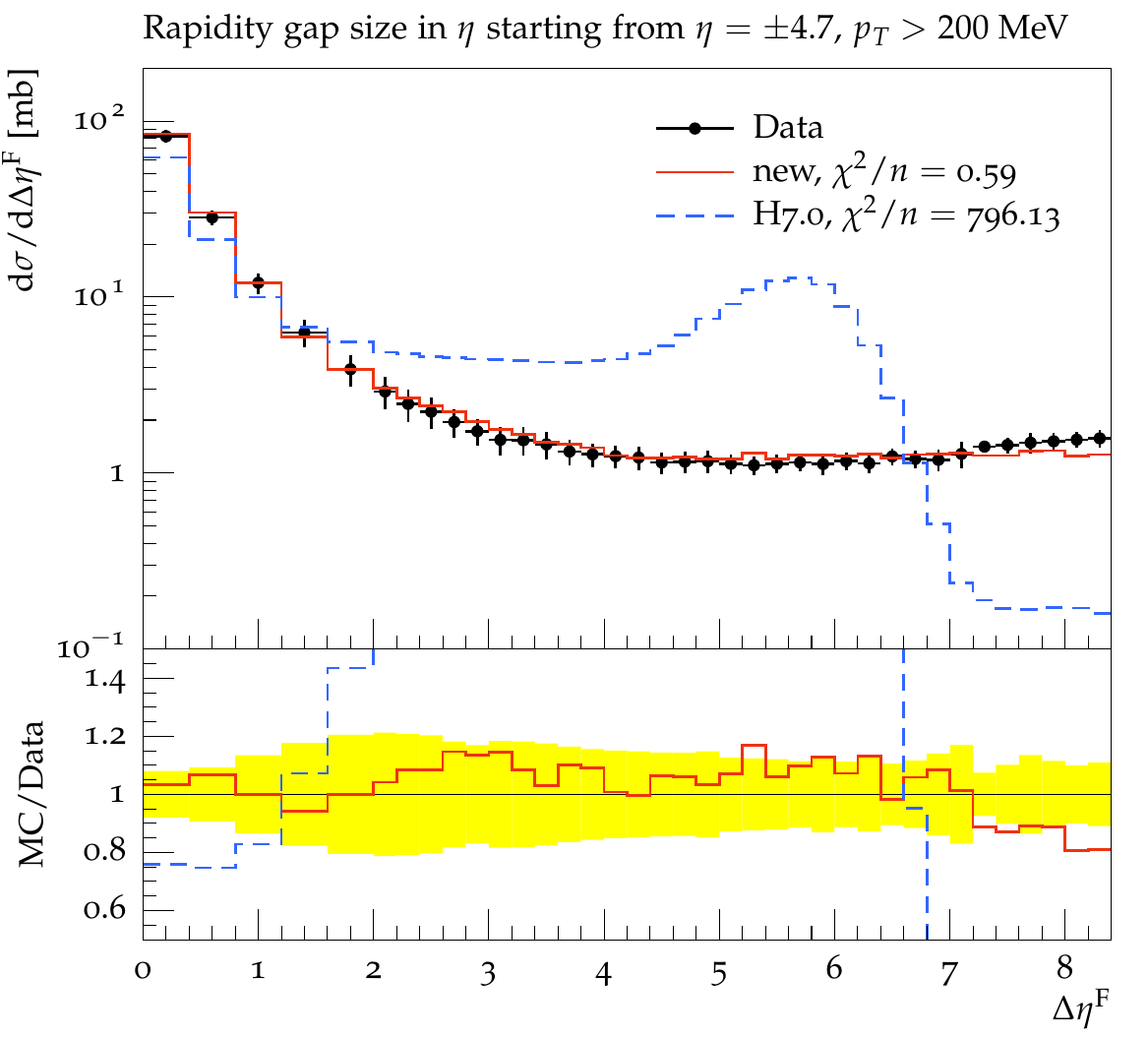}
\includegraphics[scale=0.5]{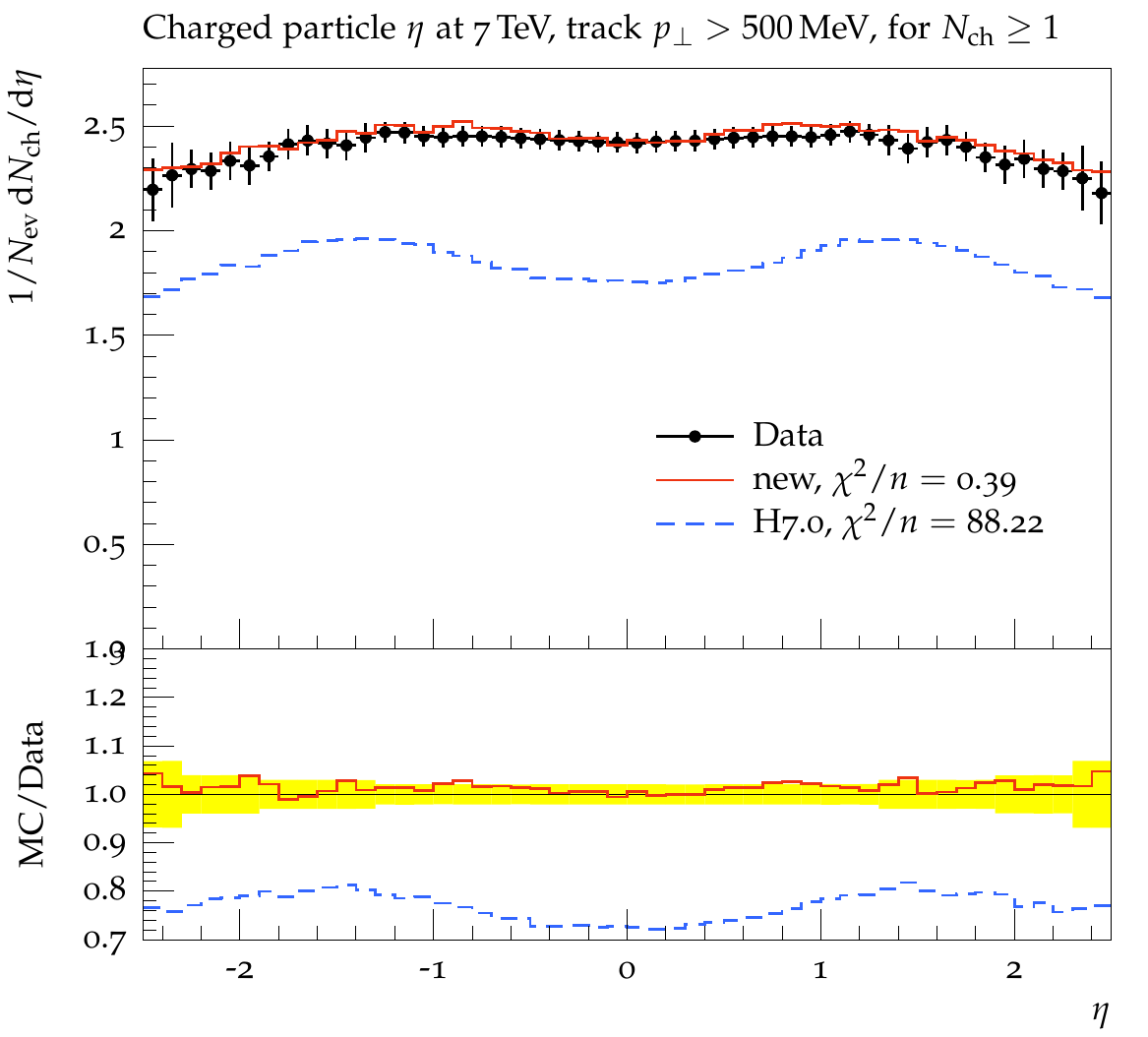}

\caption{The $\Delta \eta^F$ distribution as measured by CMS \cite{Khachatryan:2015gka} 
and the $\eta$ distribution as measured by ATLAS 
\cite{Aad:2010ac}. Shown is a comparison between the new model
for soft interactions and the old model.}
\label{fig:results}
\end{figure}

\begin{figure}[htb]
\centering
\includegraphics[scale=0.7]{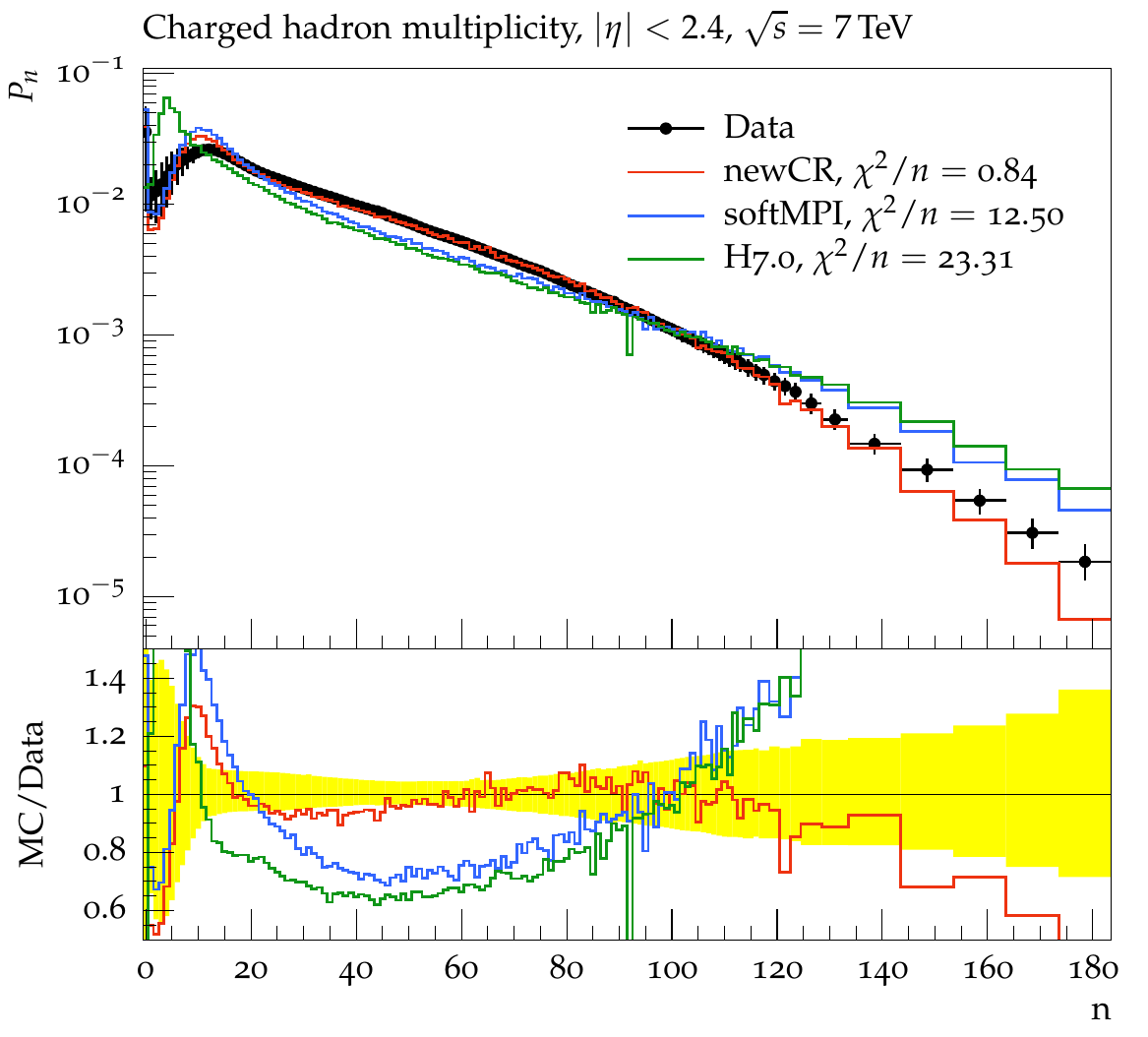}

\caption{Multiplicity distribution for the most inclusice measurement by CMS \cite{Khachatryan:2010nk}.
Shown is the new model for colour reconnection (newCR) the new soft interaction model (softMPI)
and the old model (H7).}
\label{fig:results2}
\end{figure}

In an energy independent tune the free model parameters were tuned to 
several MB observables at 900 GeV, 7 TeV and 13 TeV. 
With the tuned set of parameters we achieve a very good description of all
observables considered in the tuning. Also observables not considered in the tuning 
show a significant improvement which gives us hope that we are on the right track with
the model.
As an example we show in fig.\,\ref{fig:results} the rapidity gap (left) as measured 
by CMS \cite{Khachatryan:2015gka} and the rapidity distribution as measured by ATLAS \cite{Aad:2010ac}. 
As it can be seen the new model improves the description compared to the old model 
from Herwig 7 significantly.
In fig.\,\ref{fig:results2} we show the charged hadron multiplicity 
as measured by CMS \cite{Khachatryan:2010nk}. 
Although we note improvements in the low multiplicity region
the new model fails to describe the data correctly.
The multiplicity is mainly influenced by the invariant masses of the clusters.
The higher the invariant cluster mass, the more particles are being produced.
In the meantime we have considered a new model for colour reconnection 
which is based on a geometrical picture to form clusters out of partons which are
close in rapidity space. In fig.\,\ref{fig:results2} we also show the effect of 
this model in the multiplicity distribution.
A detailed description of this new colour reconnection model and detailed study of various
observables including flavour observables can be found in \cite{Gieseke:2017new}.

\section{Conclusion}
We have presented a new model for soft and diffractive scatterings in 
the Monte Carlo event generator Herwig. 
With the tuning to MB data a single set of parameters was found
which is able to describe all available collision energies at the LHC.
While being able to improve the majority of MB observables considered
the model fails to describe the charged mutliplicity distributions correctly.
We tackle this issue by introducing a new model for colour reconnection.



\end{document}